# Dual-Cross-Polarized GPR Measurement Method for Detection and Orientation Estimation of Shallowly Buried Elongated Object

Hai-Han Sun, Yee Hui Lee, *Senior Member*, *IEEE*, Wenhao Luo, Lai Fern Ow,
Mohamed Lokman Mohd Yusof, and Abdulkadir C. Yucel, *Senior Member*, *IEEE*

*Abstract*—Detecting a shallowly buried and elongated object and estimating its orientation using a commonly adopted co-polarized GPR system is challenging due to the presence of strong ground clutter that masks the target reflection. A cross-polarized configuration can be used to suppress ground clutter and reveal the object reflection, but it suffers from inconsistent detection capability which significantly varies with different object orientations. To address this issue, we propose a dual-cross-polarized detection (DCPD) method which utilizes two cross-polarized antennas with a special arrangement to detect the object. The signals reflected by the object and collected by the two antennas are combined in a rotationally invariant manner to ensure both effective ground clutter suppression and consistent detection irrespective of the object orientation. In addition, we present a dual-cross-polarized orientation estimation (DCPOE) algorithm to estimate the object orientation from the two cross-polarized data. The proposed DCPOE algorithm is less affected by environmental noise and performs robust and accurate azimuth angle estimation. The effectiveness of the proposed techniques on the detection and orientation estimation and their advantages over the existing method have been demonstrated using experimental data. Comparison results show that the maximum and average errors are 22.3° and 10.9° for the Alford rotation algorithm, while those are 4.9° and 1.8° for the proposed DCPOE algorithm in the demonstrated shallowly buried object cases. The proposed techniques can be unified in a framework to facilitate the investigation and mapping of shallowly buried and elongated targets.

## I. INTRODUCTION

G ROUND-PENETRATING radar (GPR) is a widely-adopted non-destructive tool for subsurface sensing and imaging [1]-[3]. One of the frequently encountered targets in GPR applications is the elongated object such as the rebar, pipe, and tree root [4], [5]. Detecting their positions and extracting their orientation information from the radargram facilitate the interpretation and classification of targets and the mapping of the subsurface utilities [6], [7]. However, for shallowly buried and elongated targets, achieving accurate detection and orientation estimation is challenging. To be specific, the reflection of these shallowly buried targets arrives almost simultaneously with the ground surface reflection. In a conventional co-polarized antenna configuration, the strong surface reflection may obscure the object's reflection [8]. In this case, it is troublesome to distinguish the object's reflection signature and to estimate the object's orientation. Therefore, new methods should be investigated to improve the detection and the orientation estimation of shallowly buried and elongated objects.

Background removal techniques such as the mean subtraction [8]-[11], the singular value decomposition (SVD) algorithm [12], [13], and the entropy-based time-gating method [14], [15] have been used to reduce the ground clutter in co-polarized GPR radargram and facilitate the detection of shallowly buried objects. Although these techniques can eliminate a major portion of clutter, their performance is subjected to surface roughness and soil conditions. Besides, the signal strength and shape of the hyperbola corresponding to target reflection can be affected by the background removal operation, especially when the object is at shallow depths. This is undesirable as they contain meaningful information for the data interpretation.

Unlike the widely used co-polarized configuration, the cross-polarized configuration is less sensitive to the surface reflection, which contains substantially the same polarized signal as the incident field, but it is sensitive to the target reflection containing depolarized signal [16]. The elongated object depolarizes the incident wave based on its orientation with respect to the antenna orientation, thus can be detected with the cross-polarized configuration. Field measurements have demonstrated that the cross-polarized antennas reduce coupling between the transmitter (TX) and receiver (RX) and the ground clutter due to polarization mismatch [16]-[21], which greatly facilitates the detection of shallowly buried and elongated targets. However, the detection capability of a cross-polarized configuration for an elongated object depends on the object's orientation. This configuration cannot detect a directional object that is parallel or orthogonal to the polarization of TX and RX [21]. *Therefore, further investigation is needed to make full use of the advantage of the cross-polarized configuration in clutter reduction while achieving a consistent detection of randomly oriented directional targets, which is a focus of this work.*

For the orientation estimation of an elongated object, several methods have been developed to extract the orientation from a GPR B-scan. Based on the reflection pattern diversity of an elongated object at different orientations, a mathematic model was established to retrieve the object orientation from its characteristic reflection curve [7]. Based on the polarization-sensitive characteristic of the elongated object, the Alford rotation algorithm was developed to extract the object direction from multi-polarimetric scattering components [22]-[26]. The estimation accuracy of these algorithms heavily relies on the clearly distinguishable reflection pattern or the accuracy of the received scattering matrix. Therefore, they are more suitable for scenarios where the object reflection can be separated from the ground clutter [7], [23]-[26]. When encountering shallowly buried objects whose reflection is interfered with ground clutter in co-polarized data, it is difficult for these algorithms to maintain high accuracy in orientation estimation.

The issues of the existence of large background clutter and the orientation-dependent detection capability when using co-linearly-polarized probes to detect shallowly buried and elongated objects also appear in the detection and characterization of tight surface-breaking cracks in metals [27]-[31]. The crack in metal, which is a complementary structure of an elongated object, scatters maximally when the polarization of the incident electromagnetic (EM) waves is perpendicular to the length of the crack. Circularly polarized antennas [30] and radially polarized antennas [31] have been implemented to perform orientation-independent detection of cracks. Algorithms based on quad-circularly-polarized synthetic aperture radar imaging have been developed for the crack detection and orientation estimation [30].

In this paper, we leverage the advantage of the cross-polarized configuration in clutter reduction and present a dual-cross-polarized detection (DCPD) method to perform clutter-mitigated orientation-independent detection of shallowly buried and elongated objects. In the method, two cross-polarized configurations with a special arrangement are used to detect a shallowly buried and elongated object. The signals received by the two configurations are combined in a rotationally invariant manner, which guarantees a consistent object detection regardless of the object orientation. In addition, a dual-cross-polarized orientation estimation (DCPOE) algorithm is developed to retrieve the object orientation information from the two cross-polarized signals. The effectiveness of the proposed techniques is validated on measured data. The DCPD method achieves consistent object detection for shallowly buried and elongated objects at different azimuth angles with similar reflection intensity. The combined cross-polarized data also produces B-scans with clear object reflection and little environmental clutter. The DCPOE algorithm automatically and accurately estimates the orientation angle with an average error of 1.8° and a maximum error of 4.9° in the demonstrated shallowly buried object cases.

## II. METHODOLOGY

In this section, we first demonstrate the advantages of cross-polarized configuration over the co-polarized configuration for the detection of shallowly buried and elongated objects. Then, we explore the information contained in the cross-polarized data and present the DCPD method and the DCPOE algorithm for the detection and orientation estimation of a shallowly buried and elongated object.

### A. The Advantage of Cross-Polarized Configuration in Detecting Shallowly Buried and Elongated Object

Field experiments with a multi-polarimetric GPR system are carried out to compare the detection capabilities of different polarizations for shallowly buried and elongated targets. Fig. 1 illustrates the measurement scenario. A metal bar with a length of 0.45 m and a diameter of 0.8 cm is buried at a depth of 3 cm in sandy soil. It is oriented at an azimuth angle of 40° with respect to the survey line. A multi-polarimetric GPR system is employed to detect the target along the survey line. In the GPR system, the dual-polarized Vivaldi antenna described in [32] is used as the transmitter (TX) and receiver (RX) in a monostatic setup. The antenna specifications are listed in Table I. It operates across a broad bandwidth from 0.4 GHz to 4.0 GHz to provide a high detection resolution. The antenna has two ports, i.e., Ports H and V, to transmit/receive horizontally and vertically polarized signals, respectively. The two ports of antennas are connected to a vector network analyzer (Keysight VNA P5008A), forming a stepped-frequency GPR system. In the experiment, the antenna is placed 2 cm above the soil surface. The output power of VNA is 10 dBm. The GPR system records 801 points of multi-polarimetric data from 0.4 GHz to 4.0 GHz in the frequency domain. The collected data are transformed to time domain via inverse Fourier transform for further analysis.

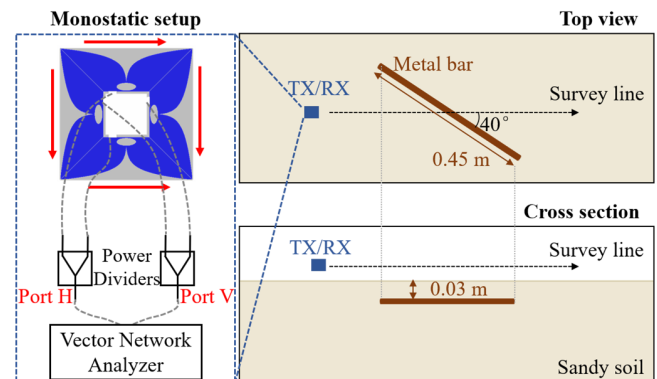

Fig. 1. The illustration of the field experiment scenario. The dual-polarized Vivaldi antenna described in [32] is used as the transmitter (TX) and receiver (RX) in a multi-polarimetric monostatic GPR system. Ports H and V excite horizontally and vertically polarized radiation, respectively. The antenna scans along a survey line to detect a shallowly buried metal bar which is oriented 40° with respect to the line. Multi-polarimetric scattering components including $S_{HH}$, $S_{HV}$, $S_{VH}$, and $S_{VV}$ are collected.

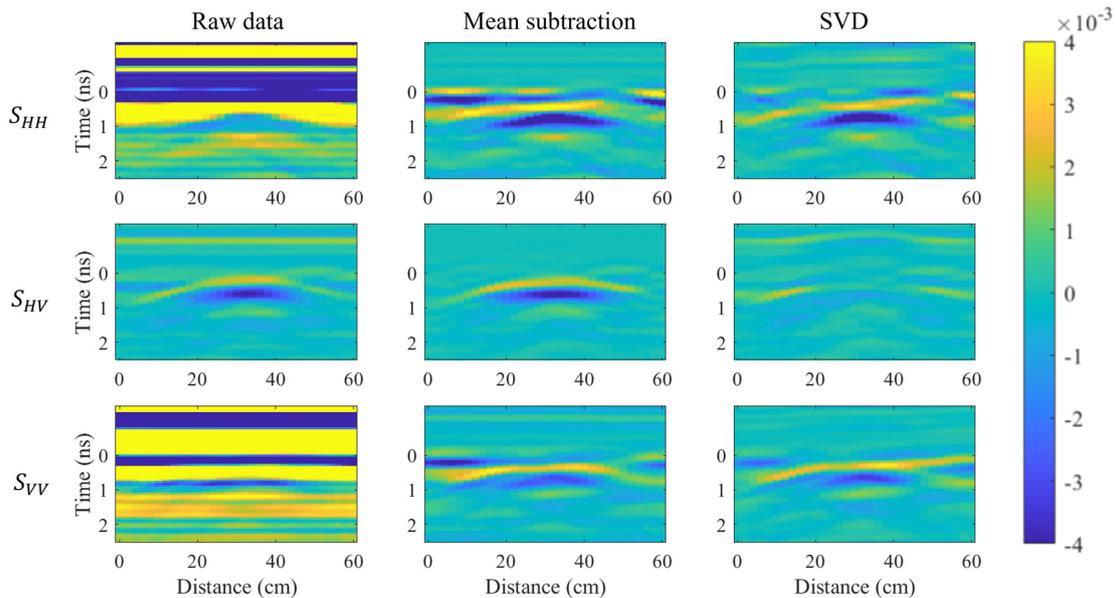

Fig. 2. Comparison of B-scans produced by the scattering components $S_{HH}$, $S_{HV}$, and $S_{VV}$ in the detection of a shallowly buried metal bar. From left to right are the raw data, data after mean subtraction, and data after SVD. **Raw data**: in the co-polarized components $S_{HH}$ and $S_{VV}$, the strong antenna and ground clutter disguise the bar's reflection, whereas in the cross-polarized component $S_{HV}$, the clutter is mitigated and the bar's reflection is clearly observed. **Data after mean subtraction**: In $S_{HH}$ and $S_{VV}$, the antenna clutter is removed, but the ground clutter cannot be fully suppressed due to slight unevenness of the surface and soil heterogeneity, leaving some clutter that interfered with the object's reflection. In $S_{HV}$, the antenna clutter and ground clutter are further suppressed, producing a B-scan with a high signal-to-clutter ratio. **Data after SVD**: the effect of SVD in $S_{HH}$ and $S_{VV}$ is similar to that of mean subtraction, but SVD is not suitable to process the $S_{HV}$ data as the dominating singular value includes the object reflection.

The scattering matrix collected by the dual-polarized antenna at a time instant can be expressed as

$$S = \begin{bmatrix} S_{HH} & S_{HV} \\ S_{VH} & S_{VV} \end{bmatrix}, \tag{1}$$

where the first and second subscripts in the matrix components represent the RX and TX ports, respectively. In the scattering matrix, $S_{HV}$ and $S_{VH}$ are equal in value based on the reciprocal principle in monostatic systems. Therefore, three different scattering components $S_{HH}$, $S_{HV}$, and $S_{VV}$ are obtained in the detection process. The original and post-processed GPR B-scans of these components are shown in Fig. 2. The post-processing techniques are the mean subtraction and the SVD [3]. The background clutter in the mean subtraction is the average trace of a B-scan collected from the site without the metal bar. The SVD is implemented by removing the largest singular value from the raw data.

As shown in Fig. 2, the raw co-polarized components $S_{HH}$ and $S_{VV}$ are dominated by the antennas' direct coupling (self-reflection in this monostatic case) and the ground surface reflection. These background clutters disguise the reflected signal from the shallowly buried metal bar. The cross-polarized component $S_{HV}$, on the contrary, shows significantly reduced background clutters and clearly reveals the hyperbola obtained by the reflections of the subsurface bar. This is because the polarization mismatch between the TX and the RX reduces the antennas' coupling, and the cross-polarized configuration is insensitive to the reflection from the ground surface.

After processing raw data by the mean subtraction or SVD technique, in $S_{HH}$ and $S_{VV}$, the clutter due to the antennas' coupling is largely removed as it is relatively constant throughout the measurement. However, the ground clutter cannot be fully removed due to slight unevenness of the surface and the soil heterogeneity, leaving some clutter that interfered with the object's reflection and distorted the hyperbola of the reflection by the target. For the cross-polarized components $S_{HV}$, the mean subtraction helps to further suppress the weak background clutter, producing an image with a high signal-to-clutter ratio (SCR). We should note that the commonly adopted SVD is not suitable to process $S_{HV}$ data as the dominating singular value is related to the object reflection. Removing the largest singular value removes the object reflection. The SCR in the $S_{HH}$, $S_{HV}$, and $S_{VV}$ data after mean subtraction, defined as the ratio of the maximum signal amplitude and the maximum clutter amplitude, are 0.98, 5.78, and 0.80, respectively. The cross-polarized data achieve significantly higher SCR compared with the co-polarized data.

The experimental results demonstrate that the cross-polarized data reveal the reflection from the shallowly buried and elongated object with a much higher SCR and clearer hyperbola when the object is oriented at an oblique angle to the incident EM field. The high SCR and clear reflection shape facilitate the detection of the elongated object and enable further data interpretation, such as the estimation of the object orientation. However, the capability of cross-polarization in detecting the elongated object is highly related to the object orientation with respect to the TX/RX. The cross-polarization cannot detect the object which is oriented parallel or perpendicular to the incident EM waves [21]. The orientation-dependent detection capability constrains the utilization of cross-polarized configuration in the detection of the shallowly buried and elongated object.

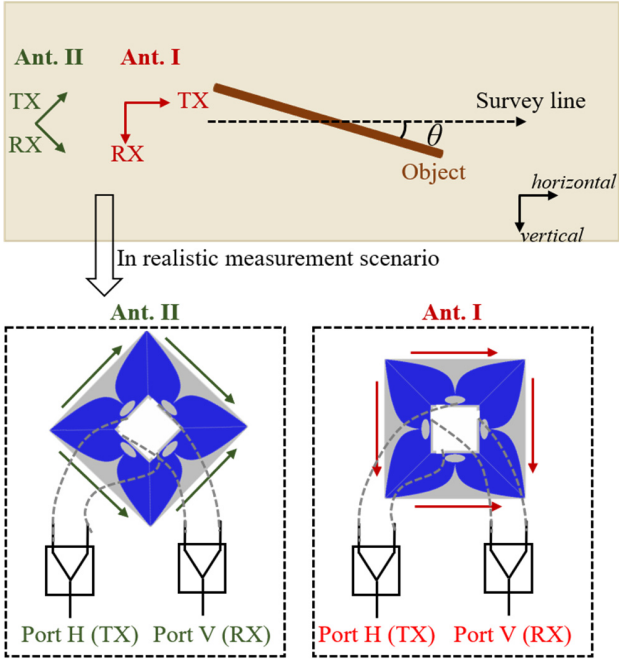

Fig. 3. Top view of the measurement scenario for detecting a shallowly buried and elongated object with an azimuth orientation angle $\theta$. Antennas I and II operate separately to scan the same survey line.

### B. The Dual-Cross-Polarized Detection (DCPD) Method

To address the limitation of the orientation-dependent detection capability of the cross-polarized configuration while leveraging its capability in clutter suppression, we propose a DCPD method to realize a consistent detection of the shallowly buried and elongated object regardless of their orientations.

The DCPD method includes a new measurement setup and a data combination mechanism. The measurement setup is shown in Fig. 3. Two cross-polarized antennas with different orientations are employed for object detection. The direction of the orthogonal polarizations in antenna I is parallel to the horizontal-vertical ($h$-$v$) coordinate. Antenna II is configured by rotating antenna I by 45° counterclockwise about its center. The elongated object is oriented at $\theta$ with respect to the survey line in the clockwise direction. The two antennas operate separately to scan the same survey line, and their collected data at the same position are aligned and combined to guarantee the consistent detection of the shallowly buried object.

The normalized scattering matrix of a horizontally oriented thin elongated object when detected via antenna I is expressed as [33]

$$S_I(\theta = 0°) = \begin{bmatrix} 1 & 0 \\ 0 & 0 \end{bmatrix}. \quad (2)$$

By applying Alford rotation transformation [22] in the clockwise direction, the scattering matrix of the object oriented at $\theta$ and detected via antenna I is calculated as

$$S_I(\theta) = \begin{bmatrix} \cos\theta & -\sin\theta \\ \sin\theta & \cos\theta \end{bmatrix} \begin{bmatrix} 1 & 0 \\ 0 & 0 \end{bmatrix} \begin{bmatrix} \cos\theta & \sin\theta \\ -\sin\theta & \cos\theta \end{bmatrix}$$

$$= \begin{bmatrix} \cos^2\theta & \frac{1}{2}\sin 2\theta \\ \frac{1}{2}\sin 2\theta & \sin^2\theta \end{bmatrix}. \quad (3)$$

The scattering matrix of the object detected via antenna II is

$$S_{II}(\theta) = \begin{bmatrix} \cos^2(\theta + 45°) & \frac{1}{2}\cos 2\theta \\ \frac{1}{2}\cos 2\theta & \sin^2(\theta + 45°) \end{bmatrix}. \quad (4)$$

For the cross-polarized component $S_{HV}$ (equivalent to $S_{VH}$ in the monostatic setup) in the $S_I(\theta)$ and $S_{II}(\theta)$, a rotationally invariant combination of cross-polarized data ($CCP$) can be obtained as

$$CCP = \sqrt{S_{I,HV}(\theta)^2 + S_{II,HV}(\theta)^2} = \frac{1}{2}. \quad (5)$$

As can be seen in (5), the sum of squares of the cross-polarized signals received by antennas I and II is constant regardless of the object orientation. Therefore, by applying the proposed dual-cross-polarized antenna setup and calculating the $CCP$, the proposed DCPD method guarantees consistent detection of a randomly oriented and elongated object, which solves the limitation of orientation-dependent detection capability of the cross-polarized configuration.

### C. The Dual-Cross-Polarization Orientation Estimation (DCPOE) Algorithm

Apart from the rotationally invariant relationship between $S_{I,HV}(\theta)$ and $S_{II,HV}(\theta)$, another important relationship between them is directly related to the object orientation angle, which is expressed as

$$S_{I,HV}(\theta)/S_{II,HV}(\theta) = \tan 2\theta. \quad (6)$$

From (6), the orientation of the elongated object can be extracted as

$$\theta_{cal} = \frac{1}{2}\tan^{-1}\left(\frac{S_{I,HV}(\theta)}{S_{II,HV}(\theta)}\right). \quad (7)$$

We should note that $S_{I,HV}(\theta)$ and $S_{II,HV}(\theta)$ in (7) are the maximum amplitudes of the object reflection, whereas in real cases, the reflected signal occupies a time span and has both positive and negative parts. Using dominating reflection amplitudes to calculate angles then averaging all the angles as the result helps to reduce the interference of noise and guarantee a more stable orientation estimation compared with only using the maximum values for calculation. However, the calculated orientation angle $\theta_{cal}$ in (7) is in the range of $[-45°, 45°)$. It can be shift to $[0°, 90°)$ by adding a period value of 90° to the negative angles, but $\theta_{cal}$ still has an ambiguity of 90° within the range $[0°, 180°)$. Fig. 4 shows the $S_{I,HV}(\theta)$ and $S_{II,HV}(\theta)$, together with the $\theta_{cal}$ at different orientation angles $\theta_{real}$ to illustrate the ambiguity. Here, we consider a target with a relative permittivity larger than the surrounding environment such as the metal bar. The reflection coefficient in this case is negative so a negative sign is applied to the scattering matrix shown in Eqs. (2)-(4). The $S_{I,HV}(\theta)$ and $S_{II,HV}(\theta)$ can be employed to eliminate the ambiguity, which is described as follows.

For the ease of description, we divide $\theta_{real}$ from 0° to 180° into four regions, A, B, A', and B', as shown in Fig. 4. $\theta_{real}$ in region A from A' and B from B' cannot be differentiated in Eq. (7) due to 90° ambiguity. There are two rules to remove the ambiguity and transform $\theta_{real}$ to the range of [0°, 180°):

- **Rule 1**: The difference between A and A' is the sign of the $S_{I,HV}(\theta)$ and $S_{II,HV}(\theta)$. If $\theta_{cal} \in [0°, 45°)$, and $S_{I,HV}(\theta)$ and $S_{II,HV}(\theta)$ are positive numbers, $\theta_{cal}$ should be added by 90° to fall in the region A'. However, when $\theta_{cal}$ is close to 0° and 45°, $S_{I,HV}(\theta)$ or $S_{II,HV}(\theta)$ is close to zero, which can be easily affected by environmental noise thus leads to inaccurate classification. Therefore, we revise this rule as if $\theta_{cal} \in [0°, 45°)$, and the number with a larger absolute value between $S_{I,HV}(\theta)$ and $S_{II,HV}(\theta)$ is a positive number, $\theta_{cal} = \theta_{cal} + 90°$.

- **Rule 2**: The difference between B and B' is the relative value of $S_{I,HV}(\theta)$ and $S_{II,HV}(\theta)$. If $\theta_{cal} \in [45°, 90°)$, and $S_{I,HV}(\theta) > S_{II,HV}(\theta)$, $\theta_{cal}$ should be added by 90° and falls in the region B'.

The pseudocode to calculate $\theta_{cal}$ from B-scan data collected by Antennas I and II and eliminate the 90° ambiguity is presented in Algorithm 1. Here, we use $CCP >$ threshold to select dominating reflection values for orientation estimation.

---

**Algorithm 1** Dual-cross-polarized orientation estimation (DCPOE) algorithm

**Input:** $S1$: cross-polarized B-scan matrix of antenna $I$, $S2$: cross-polarized B-scan matrix of antenna $II$. $S1$ and $S2$ have matrix size of [m,n], and their maximum amplitudes are $Sm1$ and $Sm2$, respectively. $threshold$: threshold to select dominating values for angle estimation

**Output:** $theta$: orientation angle of the elongated object
1: $CCP \leftarrow sqrt(S1 .* S1 + S2 .* S2)$
2: $angle \leftarrow zeros[m, n]$
3: $sum \leftarrow 0$
4: $number \leftarrow 0$
5: **for** $i = 1$ to m **do**
6:   **for** $j = 1$ to n **do**
7:     **if** $CCP[i,j] > threshold$ **then**
8:       $angle[i,j] \leftarrow arctan(S1[i,j]/S2[i,j])/2/\pi * 180$
9:       **if** $angle[i,j] < 0$ **then**
10:          $angle[i,j] \leftarrow angle[i,j] + 90$
11:       **end if**
12:       $sum \leftarrow sum + angle[i,j]$
13:       $number \leftarrow number + 1$
14:     **end if**
15:   **end for**
16: **end for**
17: $theta \leftarrow sum/number$
18: **if** $theta < 45$ **then**
19:   **if** $|Sm1| > |Sm2|$ **then**
20:     $MaxValue \leftarrow Sm1$
21:   **else**
22:     $MaxValue \leftarrow Sm2$
23:   **end if**
24:   **if** $MaxValue > 0$ **then**
25:     $theta \leftarrow theta + 90$
26:   **end if**
27: **else**
28:   **if** $Sm1 > Sm2$ **then**
29:     $theta \leftarrow theta + 90$
30:   **end if**
31: **end if**

---

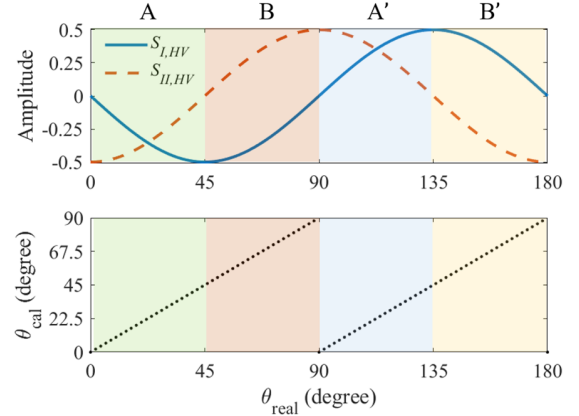

Fig. 4. The maximum amplitude of $S_{I,HV}(\theta)$ and $S_{II,HV}(\theta)$, together with $\theta_{cal}$ at different orientation angles $\theta_{real}$. 90° ambiguity can be observed in $\theta_{cal}$.

If the relative permittivity of the object is smaller than the surrounding environment, the reflection coefficient is positive, and $S_{I,HV}(\theta)$ and $S_{II,HV}(\theta)$ are 180° out-of-phase with the case shown in Fig. 4. In this case, the two rules for 90° ambiguity elimination should be revised as:

- **Rule 1**: if $\theta_{cal} \in [0°, 45°)$, and the number with a larger absolute value between $S_{I,HV}(\theta)$ and $S_{II,HV}(\theta)$ is a negative number, $\theta_{cal} = \theta_{cal} + 90°$.

- **Rule 2**: if $\theta_{cal} \in [45°, 90°)$, and $S_{I,HV}(\theta) < S_{II,HV}(\theta)$, $\theta_{cal} = \theta_{cal} + 90°$.

The pseudocode should be revised in lines 24 and 28 accordingly. The application of the DCPOE requires a prior knowledge of the comparison of permittivity values of the targets and the medium. In our applications, the permittivity of targets is greater than the permittivity of soil.

The validity of the proposed DCPD method and DCPOE algorithm have been examined using experimental data. The implementation details and the results are described in Section III.

It is noted that the DCPD method and the DCPOE algorithm are developed by only collecting the cross-polarized signals. They leverage the ground clutter suppression capability of the cross-polarized configuration and facilitate the detection and characterization of shallowly buried and elongated objects. They could be adopted by conventional GPR systems that only obtain one polarization component at a time. If a fully-polarimetric GPR system is available, similar orientation-independent detection capability can be achieved by using quad-circularly-polarized measurement or using quad-linearly-polarized measurement and converting the obtained S-parameters to the quad-circularly-polarized S-parameters [30]. In addition, using the converted quad-circularly-polarized S-parameters, a quad-pol algorithm was introduced in [30] to address the issue of 90° ambiguity in the orientation estimation. The algorithm in [30] was developed for the orientation estimation of cracks in metals. The performance of the algorithm for estimating orientations of shallowly buried objects could be slightly affected by the rough and uneven soil surface in GPR applications.

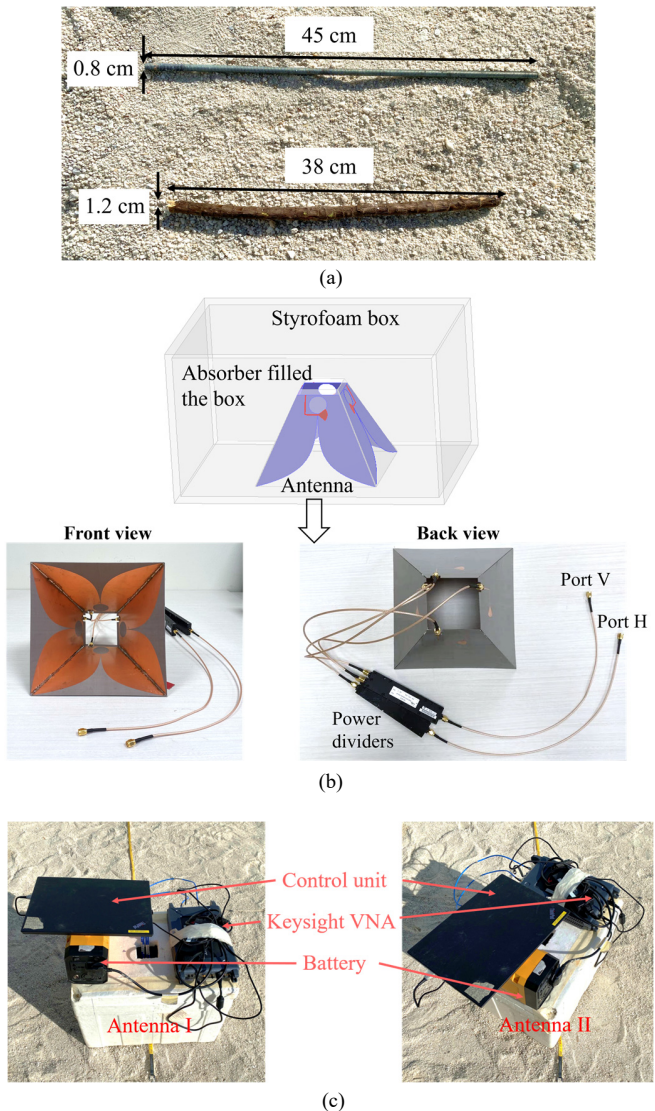

Fig. 5. The testing scenarios. (a) A metal bar and a tree root that used as elongated targets in the experiments. (b) Sketch of the antenna box and pictures of the antenna in [32] used for measurement. (c) The antenna box is placed as antennas I and II to scan the same trace. Antenna II is achieved by rotating antenna I by 45° counterclockwise about its center.

## III. Experimental Validation

Field experiments are carried out to validate the proposed techniques. The testing scenarios are shown in Fig. 5. The elongated subsurface targets in the experiment are a metal bar and a tree root, which are commonly encountered engineering and environmental objects, respectively. As shown in Fig. 5(a), the metal bar has a length of 45 cm and a diameter of 0.8 cm, and the tree root has a length of 38 cm with a diameter of 1.2 cm. They are buried in sandy soil at a depth of around 2-3 cm, and are oriented at 10 different angles 0°, 20°, 40°, 60°, 80°, 90°, 100°, 120°, 140°, and 160°. The relative permittivity of the sandy soil and the tree root are around 3 and 25, respectively, which are measured using the Keysight N1501A dielectric probe.

The measurement scenario is the same as the one illustrated in Fig. 3. The antenna in the experiment is the dual-polarized Vivaldi antenna described in [32], as shown in Fig. 5(b). It is sealed in a Styrofoam box and surrounded by absorbers to reduce environmental noise. The antenna box is first placed following the polarization direction of antenna I to scan the survey line, and then rotate 45° counterclockwise as antenna II to scan the same line, as shown in Fig. 5(c). The system setup is the same as the monostatic system described in Section II.A. The cross-polarized data ($S_{21}$) from the two measurements are aligned in the survey position and are used to implement the proposed techniques. All the experimental results are presented as follows.

### A. Experimental Results with Metal Bar

First, we examine the capability of our proposed DCPD method in detecting the shallowly buried object. Fig. 6 shows six sets of the B-scans of cross-polarized data collected by antennas I and II, and their combined image using $CCP$ values at different orientation angles 0°, 40°, 80°, 90°, 120°, and 160°. Mean subtraction is used to pre-process the data.

As shown in Fig. 6, the detection capability of antennas I and II significantly varies in different orientation cases. Relying on only one antenna cannot guarantee the detection of the elongated object (for example, object reflection is indistinguishable in B-scans of Antenna I at 0° and 90° and of Antenna II at 40°). Using the proposed combination mechanism, $CCP$ combines the information carried in two cross-polarized data and produces B-scans with clearly distinguished object reflection and little ground clutter regardless of the object orientation. The maximum $CCP$ values in all cases are listed in Table I. The values in different orientation cases are very close to each other. The small discrepancies between them can be caused by the slight positional misalignment between the two cross-polarized data, and the variation of the object depth and soil's relative permittivity in different cases. The $CCP$ values listed in Table II together with the $CCP$ images shown in Fig. 6 prove that the proposed DCPD method is capable of conducting a consistent detection of the shallowly buried and elongated target regardless of its orientation and producing a clearer detection image with little environmental clutter.

Then, the orientation estimation accuracy of the proposed DCPOE algorithm is investigated with the experimental data. As complex reflection coefficients are obtained, the real part of the ratio of the two cross-polarized data is taken before applying the arctan function in (7). We should note that if the reflection amplitudes are small, the orientation estimation can be easily affected by the noise in the data. Therefore, we use normalized $CCP$ values $\overline{CCP} > TH$ as a threshold for the selection of the amplitudes to implement the algorithm. The estimation accuracy evaluated by the mean absolute error in the 10 metal bar cases with different $TH$ values is shown in Fig. 7. A larger $TH$ selects more dominant reflection values of the metal bar to perform orientation estimation, which helps to reduce the interference of environmental noise and improve the estimation accuracy. The variation of estimation error is very small for $TH \geq 0.8$ in both the metal bar and the tree root cases, therefore we choose $TH = 0.8$ to perform a stable and accurate orientation estimation.

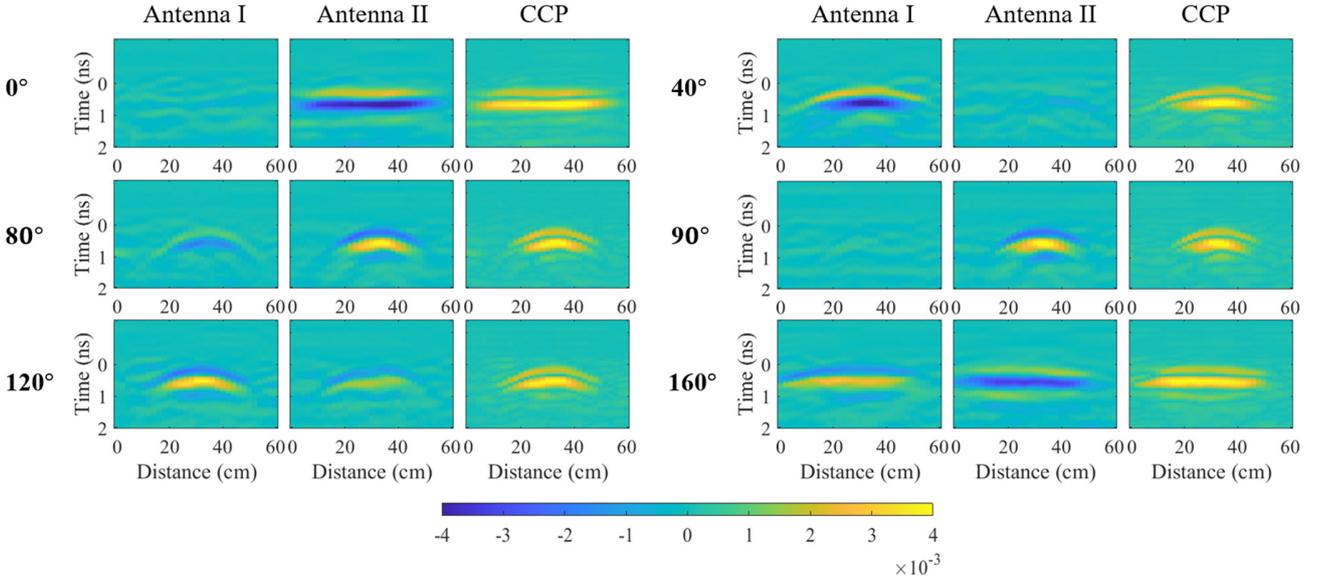

Fig. 6. Six sets of cross-polarized post-processed B-scans collected from antennas I and II, and the combined B-scans using the $CCP$ values for the metal bar oriented at 0°, 40°, 80°, 120°, and 160°. Antennas I and II have significantly different responses at different orientation cases. Using only one antenna cannot guarantee the detection of the object. For example, antenna I at 0° and 90°, and antenna II at 40° result in indistinguishable object reflection. However, the proposed algorithm produces $CCP$ images with very clear object reflection and little ground clutter regardless of the object orientation.

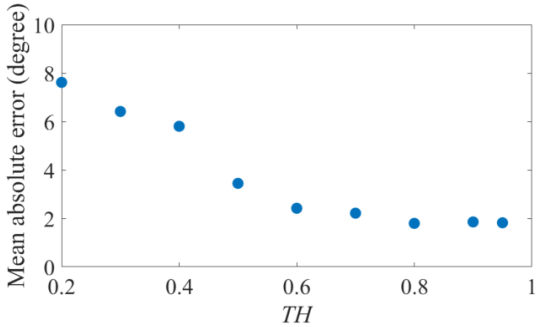

Fig. 7. The mean absolute error of orientation estimation with different threshold values ($\overline{CCP} > TH$) in the metal bar cases.

Fig. 8 shows the orientation estimation results for the six cases presented in Fig. 6. The continuous lines in the background are the A-scans with $CCP$ values. The estimated angles are presented in color scale. As observed, with the threshold applied, the algorithm automatically selects the regions with strong object reflection for angle estimation, which is less prone to noise. The estimated angles in each subfigure are shown as almost identical color, proving the stability of orientation estimation process. The average estimated angle in the plot ($\theta_{cal}$) and the real angle ($\theta_{real}$) as marked in the figures. They are very close in value, demonstrating the estimation accuracy of the DCPOE algorithm. The estimation results in all 10 cases are listed in Table III. The maximum and average angle estimation errors are 4.9° and 1.8°, respectively. The experimental results prove the validity of the DCPOE algorithm in stable and accurate orientation estimation.

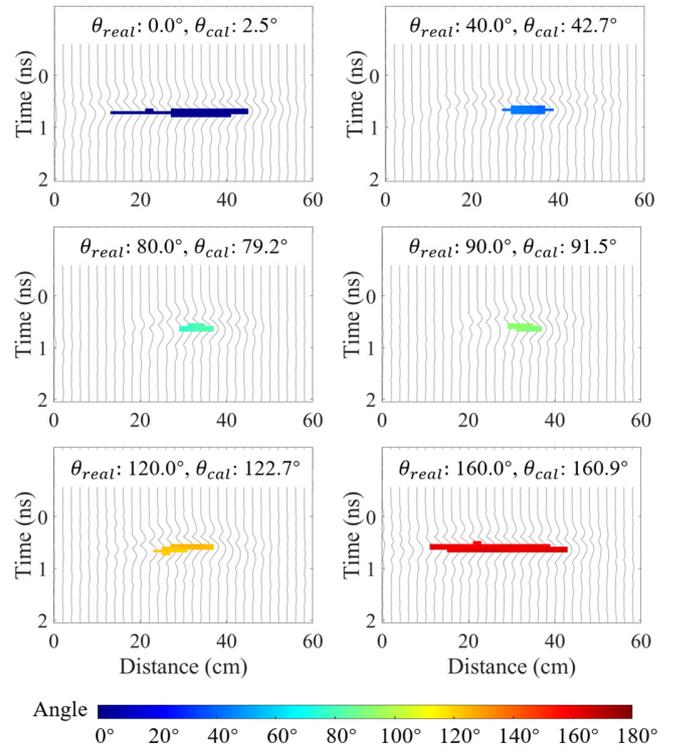

Fig. 8. The angle estimation results of our DCPOE algorithm for the metal bar oriented at angles 0°, 40°, 80°, 90°, 120°, and 160°. Different angles are represented by different colors. $\overline{CCP} > 0.8$ is applied as a threshold to guarantee a stable angle estimation.

TABLE II
COMBINED VALUES ($CCP$) OF THE TWO CROSS-POLARIZED DATA IN THE
METAL BAR CASES

| $\theta_{real}$ | 0° | 20° | 40° | 60° | 80° |
|---|---|---|---|---|---|
| $CCP$ values | 0.0045 | 0.0042 | 0.0047 | 0.0047 | 0.0046 |
| $\theta_{real}$ | 90° | 100° | 120° | 140° | 160° |
| $CCP$ values | 0.0047 | 0.0043 | 0.0042 | 0.0047 | 0.0043 |

TABLE III
ORIENTATION ESTIMATION RESULTS USING THE PROPOSED DCPOE
ALGORITHM IN THE METAL BAR CASES

| $\theta_{real}$ | 0° | 20° | 40° | 60° | 80° |
|---|---|---|---|---|---|
| $\theta_{cal}$ | 2.5° | 21.9° | 42.7° | 60.0° | 79.2° |
| $\theta_{real}$ | 90° | 100° * | 120° | 140° | 160° |
| $\theta_{cal}$ | 91.5° | 104.9° | 122.7° | 139.9° | 160.9° |

* maximum error case.

## B. Experimental Results with Tree Root

The tree root presents an important environmental target in GPR applications. Unlike the elongated engineering targets such as the metal bars which are straight, most roots have some curvature such as the one shown in Fig. 5(a). In this experiment, we examine the detection capability of the DCPD method and the orientation estimation accuracy of the DCPOE algorithm with the shallowly buried root. The orientation of the root with a small curvature is defined by its starting and ending point.

The post-processed B-scans collected by antennas I and II, and the combined B-scans using $CCP$ values of six orientation cases are shown in Fig. 9. Similar to the findings in the metal bar cases, $CCP$ combines the information collected by antennas I and II, and presents the object reflection clearly with a high SCR. The maximum $CCP$ values in different cases are listed in Table IV. The root curvature, the slight misalignment of data, and the small variation of the environmental condition can result in small discrepancies in $CCP$ values. Nevertheless, the values are still close to each other. The results demonstrate that the DCPD method maintains its consistent detection capability for a randomly oriented root with small curvature. The dual-cross-polarized data are further processed with the DCPOE algorithm to extract the root orientation angles. The estimated results in six cases are shown in Fig. 10. Consistent color is shown in each plot, indicating a stable angle estimation process. The average estimated angles are marked in the figures, which are very close to the real angles. Table V shows the angle estimation results in the 10 cases. The DCPOE algorithm accurately estimates the angles with the maximum and average errors of 4.3° and 1.7°, respectively. The results further demonstrate that the DCPOE algorithm has high accuracy in the orientation estimation even when the elongated object has a small curvature.

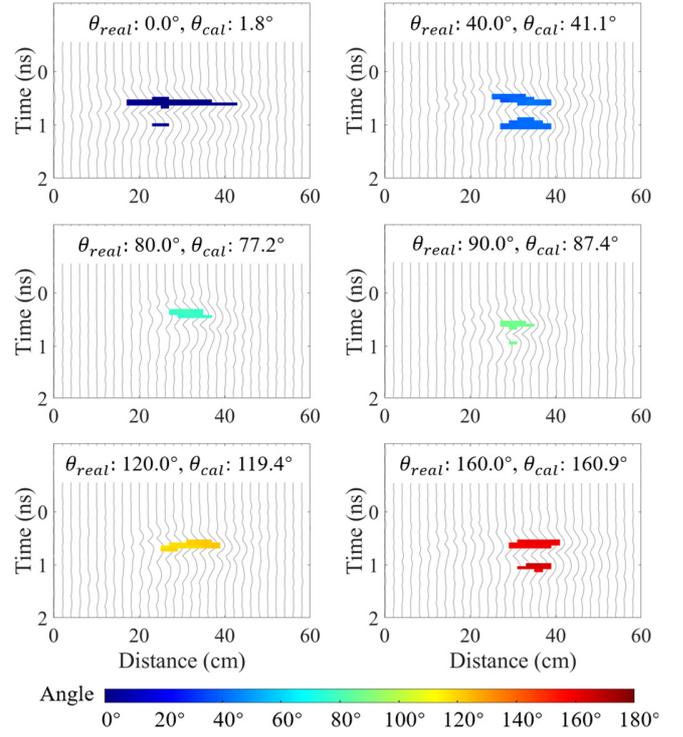

Fig. 10. The angle estimation results of the proposed DCPOE algorithm for the tree root at real angles 0°, 40°, 80°, 90°, 120°, and 160°. Different angles are represented by different colors. $\overline{CCP} > 0.8$ is applied as a threshold to guarantee a stable angle estimation.

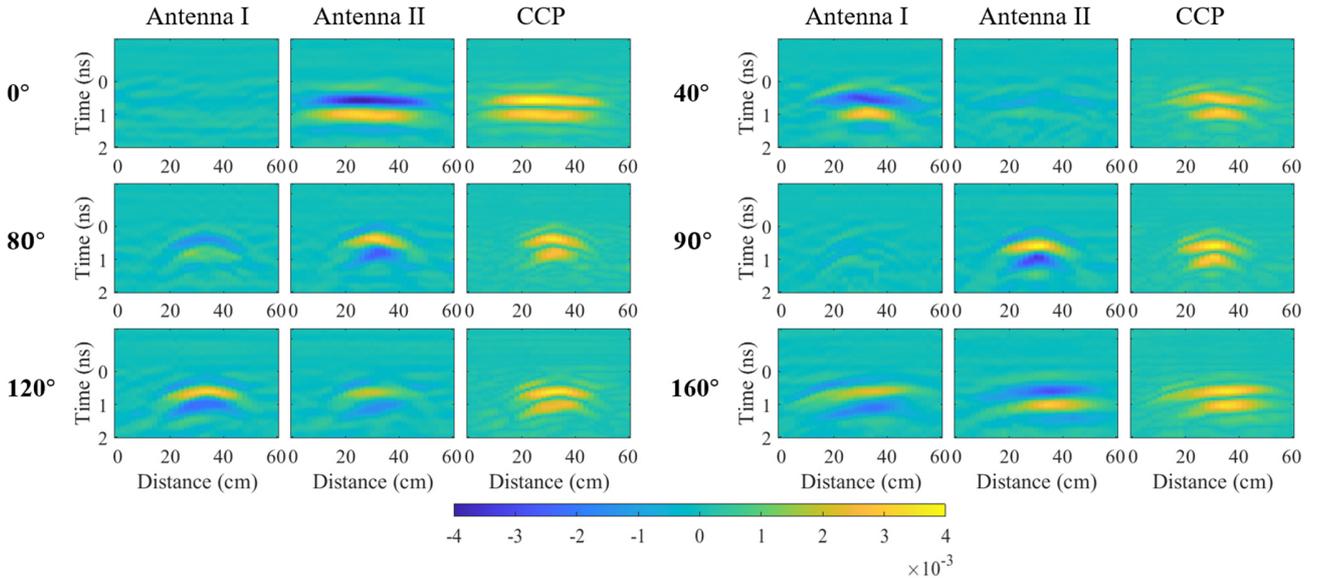

Fig. 9. Six sets of cross-polarized post-processed B-scans collected from antennas I and II, and the combined B-scans of $CCP$ values for the tree root oriented at 0°, 40°, 80°, 120°, and 160°. Similar to the results shown in Fig. 5 with the metal bar, Antennas I and II have significantly different responses in different orientation cases, whereas the proposed method produces $CCP$ images with very clear object reflection and little environmental clutter regardless of the orientation.

TABLE IV
COMBINED VALUES ($CCP$) OF THE TWO CROSS-POLARIZED DATA IN THE TREE ROOT CASES

| $\theta_{real}$ | 0° | 20° | 40° | 60° | 80° |
|---|---|---|---|---|---|
| $CCP$ values | 0.0041 | 0.0036 | 0.0038 | 0.0036 | 0.0037 |
| $\theta_{real}$ | 90° | 100° | 120° | 140° | 160° |
| $CCP$ values | 0.0042 | 0.0037 | 0.0039 | 0.0041 | 0.0038 |

TABLE V
ORIENTATION ESTIMATION RESULTS USING THE PROPOSED DCPOE ALGORITHM IN THE TREE ROOT CASES

| $\theta_{real}$ | 0° | 20° | 40° | 60° | 80° |
|---|---|---|---|---|---|
| $\theta_{cal}$ | 1.8° | 20.3° | 41.1° | 61.5° | 77.2° |
| $\theta_{real}$ | 90° | 100° | 120° | 140° * | 160° |
| $\theta_{cal}$ | 87.4° | 98.7° | 119.4° | 135.7° | 160.9° |

* maximum error case.

### C. Application of the DCPD method and the DCPOE algorithm for the Investigation of Elongated Objects at Other Depths

Although the DCPD method and the DCPOE algorithm are developed to address the issues of the interference of large ground clutter and the orientation-dependent detection capability when using co-polarized GPR systems to detect shallowly buried and elongated objects, their operating mechanism is applicable to the investigation of elongated objects at other depths. To verify that, we conduct experiments with the metal bar shown in Fig. 5(a) buried at a larger depth of 21 cm in sandy soil. The metal bar is oriented at five different angles 0°, 40°, 80°, 120°, and 160°. The calculated $CCP$ values and the estimated orientation angles using the DCPD method and the DCPOE algorithm in different cases are listed in Table VI. The B-scans using the $CCP$ values and the visualization of estimated orientation angles are shown in Fig. 11.

As shown in Table VI and Fig. 11, the $CCP$ values in different orientation cases are close to each other. The object reflection is very clear in the B-scans produced by $CCP$ values in each case, demonstrating the orientation-independent detection capability of the DCPD method. The $CCP$ values in the 21 cm cases are smaller than those in the 3 cm cases (around 0.0045). This is due to the attenuation of the electromagnetic waves with a longer traveling distance in soil. The estimated orientation angles listed in Table VI are very close to the real orientation angles. The mean absolute error in the five cases is 2.4°, proving the estimation accuracy of the DCPOE algorithm for an elongated object at a deeper location. As shown in Fig. 11, the algorithm automatically selects regions with strong object reflection for angle estimation, and it achieves stable and accurate estimation performance in each case. The results presented in Section III.A-C verify that the DCPD method and the DCPOE algorithm maintain their effectiveness in investigating elongated objects buried at different depths.

TABLE VI
COMBINED VALUES ($CCP$) AND ESTIMATED ORIENTATION ANGLES FOR THE METAL BAR AT A DEPTH OF 21 CM

| $\theta_{real}$ | 0° | 40° | 80° | 120° | 160° |
|---|---|---|---|---|---|
| $CCP$ values | 0.0028 | 0.0026 | 0.0023 | 0.0023 | 0.0026 |
| $\theta_{cal}$ | 1.9° | 41.3° | 75.4° | 122.1° | 162.2° |

### D. Comparison and Discussion

We implement the well-developed Alford rotation algorithm [25] to estimate the angles in the shallowly buried metal bar and tree root cases and compare its accuracy with our proposed DCPOE algorithm. For the angle estimation using the Alford rotation algorithm, we use the scattering matrix obtained by antenna I, and apply the mean subtraction to process the B-scans of different polarimetric components. The normalized absolute value of the trace of the scattering matrix $|S_{1,HH}(\theta) + S_{1,VV}(\theta)| > 0.5$ is used as a threshold for the selection of amplitudes [23]. This threshold is chosen after experimenting with different values to achieve the minimum estimation error. Six results in the metal bar cases are presented in Fig. 12. We note that $\theta_{cal}$ is the average estimated angle only in the manually selected object reflection region. If the average value of the angles in the entire figure is used as the result, a huge estimation error occurs due to the presence of environmental clutter.

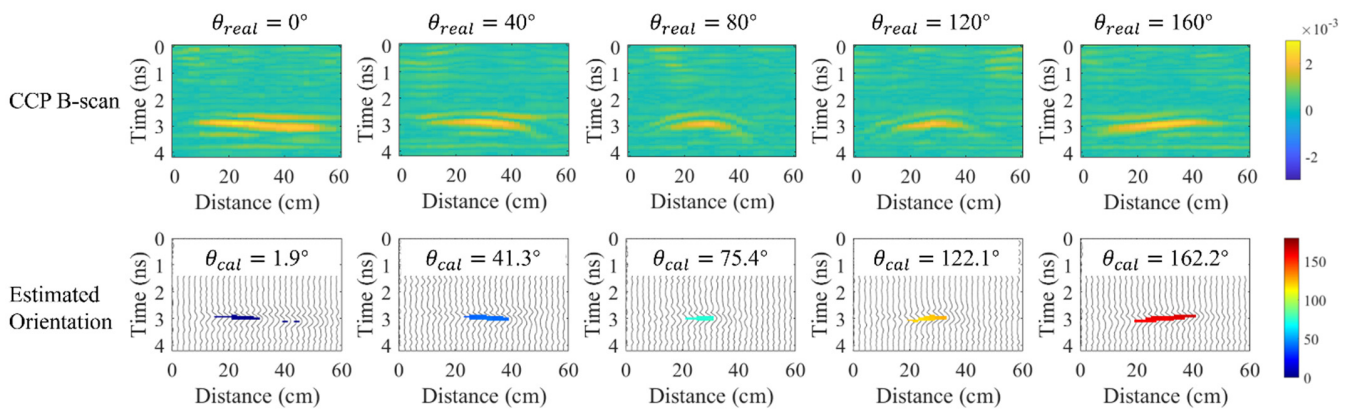

Fig. 11. The B-scans of the $CCP$ values and the visualization of estimated orientation angles when the metal bar is buried at a depth of 21 cm.

Comparing the results of Alford rotation algorithm shown in Fig. 12 with the results obtained by the DCPOE algorithm shown in Fig. 8, it is evident that: 1) Fig. 12 contains pronounced environmental clutter from the co-polarized components, which has similar or even larger strength to the target response, making the differentiation of the object difficult, such as in the 80°, 90°, and 120° cases. 2) The estimation in the object reflection region is less stable, as shown by different color in the same figure, such as in the 0° and 80° cases. This is due to the existence of environmental clutter that interfered with the target response in the co-polarized scattering components.

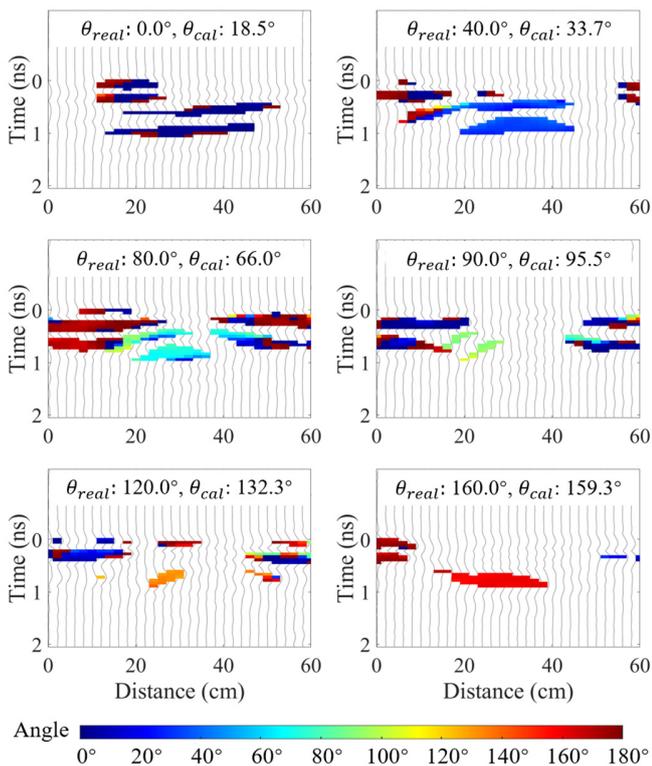

Fig. 12. The angle estimation results using Alford rotation algorithm presented in [25] for the metal bar at real angles 0°, 40°, 80°, 90°, 120°, and 160°. Different angles are represented by different colors. The background trace is A-scans with the normalized absolute value of the trace of the scattering matrix $|\text{Tr}(S_1(\theta))| = |S_{1,HH}(\theta) + S_{1,VV}(\theta)|$. The images contain pronounced environmental clutter from the co-polarized components, making the differentiation of the object reflection difficult in some cases. The angle estimation is less accurate due to the interference of environmental noise on the object response in the co-polarized scattering components.

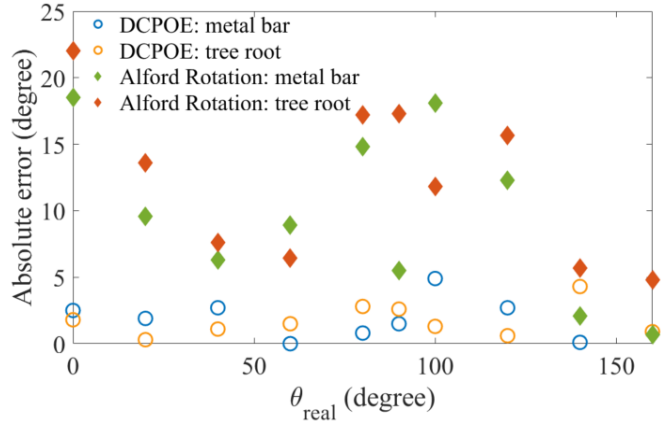

Fig. 13. Absolute angle estimation error for the DCPOE algorithm and the Alford rotation algorithm in the metal bar and tree root cases. The estimation error of the proposed DCPOE is significantly smaller than that of the Alford rotation algorithm in most cases.

The absolute angle estimation error obtained using the Alford rotation algorithm and the proposed DCPOE algorithm in all cases are shown in Fig. 13. The estimation error of the Alford rotation algorithm is larger than the proposed DCPOE algorithm in most cases. The maximum and average errors in all cases are 22.3° and 10.9° for the Alford rotation algorithm, and are 4.9° and 1.8° for the proposed DCPOE algorithm, respectively.

Compared with the Alford rotation algorithm, the DCPOE algorithm is less affected by the environmental clutter, and the estimation accuracy is significantly higher for the shallowly buried and elongated objects. The comparison results demonstrate the advantages of our proposed dual-cross-polarized measurement method in reducing ground clutter and improving orientation estimation accuracy for shallowly buried and elongated objects.

## IV. CONCLUSION

In this paper, we present a DCPD method together with a DCPOE algorithm for the investigation of shallowly buried and elongated objects. The DCPD method uses two specially arranged cross-polarized configurations for detection. The two cross-polarized signals collected by the two configurations are combined in a rotationally invariant manner, which guarantees consistent detection of elongated objects regardless of their orientations. The utilization of cross-polarized data helps to produce clearly revealed object reflection with a high SCR. The DCPOE algorithm builds the connection between the two cross-polarized signal strengths with the object orientation. It is capable of achieving robust and accurate object orientation estimation. Experimental results with a shallowly buried metal bar and a tree root have demonstrated the effectiveness of the proposed algorithms. Although the DCPD method and the DCPOE algorithm are developed to address issues in the investigation of shallowly buried and elongated objects, their working mechanism is applicable to the study of elongated objects in general, which has been verified using the experimental data with a metal bar buried at a depth of 21 cm.

The DCPD method and the DCPOE algorithm presented in this paper are developed based on the ground clutter suppression capability of the cross-polarized configuration and the depolarization mechanism of the elongated object. They can be implemented as a unified framework to perform the consistent detection of subsurface elongated objects and the accurate estimation of the objects' orientation. In addition, the promising results obtained using only the cross-polarized data in this study can provide insights into further exploration of the cross-polarized configuration in GPR applications.